\documentclass{article}

\usepackage[numbers]{natbib}

\usepackage{tabularx}
\usepackage{graphicx}

\usepackage{amsmath}  
\usepackage{amsfonts} 
\usepackage{amssymb}  
\usepackage{amsthm}   
\usepackage{hyperref} 
\usepackage{cleveref}
\theoremstyle{plain} 

\newtheorem{theorem}{Theorem}

\theoremstyle{definition}
\newtheorem{definition}{Definition}

\theoremstyle{lemma}

\theoremstyle{corollary}

\crefformat{equation}{\textnormal{Eq. #2#1#3}}
\crefrangeformat{equation}{\textnormal{Eqs. #3#1#4 - #5#2#6}}
\crefmultiformat{equation}{\textnormal{Eqs. #2#1#3}}
{ and #2#1#3}{, #2#1#3}{ and #2#1#3}
\crefformat{figure}{Fig. #2#1#3}
\crefrangeformat{figure}{Figs. #3#1#4 - #5#2#6}
\crefmultiformat{figure}{Figs. #2#1#3}
{ and #2#1#3}{, #2#1#3}{ and #2#1#3}
\crefformat{table}{Table #2#1#3}
\crefrangeformat{table}{Tables #3#1#4 - #5#2#6}
\crefmultiformat{table}{Tables #2#1#3}
{ and #2#1#3}{, #2#1#3}{ and #2#1#3}
\crefformat{appendix}{Appendix #2#1#3}
\crefrangeformat{appendix}{Appendixes #3#1#4 - #5#2#6}
\crefmultiformat{appendix}{Appendixes #2#1#3}
{ and #2#1#3}{, #2#1#3}{ and #2#1#3}
\crefformat{section}{Section #2#1#3}
\crefrangeformat{section}{Sections #3#1#4 - #5#2#6}
\crefmultiformat{section}{Sections #2#1#3}
{ and #2#1#3}{, #2#1#3}{ and #2#1#3}
\crefformat{definition}{Definition #2#1#3}
\crefrangeformat{definition}{Definitions #3#1#4 - #5#2#6}
\crefmultiformat{definition}{Definitions #2#1#3}
{ and #2#1#3}{, #2#1#3}{ and #2#1#3}
\crefformat{theorem}{Theorem #2#1#3}
\crefrangeformat{theorem}{Theorems #3#1#4 - #5#2#6}
\crefmultiformat{theorem}{Theorems #2#1#3}
{ and #2#1#3}{, #2#1#3}{ and #2#1#3}
\crefformat{lemma}{Lemma #2#1#3}
\crefrangeformat{lemma}{Lemmas #3#1#4 - #5#2#6}
\crefmultiformat{lemma}{Lemmas #2#1#3}
{ and #2#1#3}{, #2#1#3}{ and #2#1#3}
\crefformat{corollary}{Corollary #2#1#3}
\crefrangeformat{corollary}{Corollaries #3#1#4 - #5#2#6}
\crefmultiformat{corollary}{Corollaries #2#1#3}
{ and #2#1#3}{, #2#1#3}{ and #2#1#3}

\usepackage{authblk} 

\title{Stochastic Modeling of the Mean Time Between Software Failures: A review } 

\author[1]{Gabriel Pena}
\author[1]{Verónica Moreno}
\author[1,2]{Néstor Barraza}

\affil[1]{Universidad Nacional de Tres de Febrero, Departamento de Ciencia y Tecnología, Caseros, Argentina}
\affil[2]{Universidad de Buenos Aires, Facultad de Ingeniería, Ciudad Autónoma de Buenos Aires, Argentina}

\date{} 

\usepackage{tabularx} 
\usepackage{array}    
\newcolumntype{X}{>{\centering\arraybackslash}X} 

\begin{document}


\title{Stochastic Modeling of the Mean Time Between Software Failures: A review}
\label{chap1}
\maketitle

\begin{abstract}
A review of several models for the software failure detection stochastic process is performed. We focus on how the Mean Time Between Failures is obtained from those models and present a comparative study. The models considered involve some of the well known based on non-homogeneous Poisson processes and a recently proposed model based on contagion. Our main target is to evaluate the models performance when they are applied to modern software projects developed under Agile methodologies. Recently developed Software Engineering practices (TDD, Scrum, etc.), propose an iterative procedure between development, testing and debugging, introducing this way, particular challenges to Software Reliability modeling. As opposed to the non-homogeneous Poisson processes, the failure rate of the contagion model depends not just on time but also on the number of previously detected failures. All the considered models are particular cases of non-homogeneous Birth Processes, a general fundamental theory on how to obtain time to failure and time between failures expectations is also shown.	
\end{abstract}

\vspace{1em} 

\noindent \textbf{Keywords:} Stochastic model, Non-homogeneous birth process, Markov, Mean time between failures, Software reliability, Agile.

\vspace{2em} 

\section{Introduction}\label{sec1}

Modeling the stochastic software failures detection process turned an important issue in Engineering in the 70's. Many software reliability models have been proposed since then where the Software Reliability Growth Models (SRGM) based on the non-homogeneous Poisson process constitutes an important group, see for example \cite{Pham:2010:SSR:1965277} and \cite{Lyu:1996:HSR:239425}. A diffusive process was recently proposed in \cite{Deepika}. A recent survey has been published in \cite{Shiva}. During the last decades, software has become a critical component in many technological projects. New development and testing methodologies were introduced in the discipline of Software Engineering with the purpose of obtaining more reliable products with quality assurance. Those methodologies, for example Agile, which involves modern practices like TDD (Test driven development), Scrum, etc., introduce new challenges to Software Reliability modeling. Since the main characteristic in Agile procedures is that development is performed simultaneously with testing in an iterative cycle, it is a quite different situation from that when the SRGM were developed when waterfall was the prevailing software development methodology. We then draw the attention in order to review the software reliability growth concept and consider the importance of taking into account a first increasing failure rate long stage. Contagious stochastic processes have been introduced in software reliability because of that, see for example \cite{7789377}. As opposed to the non-homogeneous Poisson processes, the failure rate in models based on contagion depends not just on time, but also on the number of failures previously detected, the contagion phenomena could model this way interaction between programmers and testers, or sharing and reuse of code. The recently introduced contagion model is a modification of the Polya stochastic process, a different functional form of the failure rate was proposed in order to obtain either a positive or negative concavity in the mean number of failures curve, in order to be able to model increasing failure rate or reliability growth stages, as it is explained in \cite{barrazaanand}, \cite{penabarrazaanand} and references therein.
Mean time between failures (MTBF) and mean time to failure (MTTF) are important and useful metrics in reliability analysis, see for example \cite{barlow}. In the particular case of Software Reliability, those parameters are useful to determine the time to release, make adjustments in the development and testing processes, when evaluated after delivery: to evaluate the performance of a given release and to set up requirements for future releases. In a reliability growth case, we expect a continuously increasing MTBF, however, taking into account an important increasing failure rate first stage as explained before, we should expect a decreasing MTBF at first, the so called ''infant mortality'' stage in hardware reliability. Then, the well known bathtub curve from hardware reliability is also applicable to software reliability. We consider then that: the way this metric is obtained, its properties, how it compares between different models and its goodness of fitting real projects are important issues to be studied. We focus then in this chapter on the analysis of the MTBF metric as obtained from several models, i.e. reliability growth models and also the recently proposed model based on contagion. We show the results of comparison of their performance when applied to real projects developed under Agile practices. As to present a clear and self explained content, we expose a complete mathematical foundation about the way the metrics are obtained. Since either the contagion based model as those based on non-homogeneous Poisson processes are birth processes, we present a mathematical background section on this issue.  

\section{Mathematical background}
	
	\subsection{Birth processes}
	
	Birth processes are a particular case of branching processes and continuous time Markov chains, they can also be considered as one of the various possible generalizations of Poisson processes. As shown for example in \cite{Feller1} a series of proposed axioms are translated into a set of ordinary differential equations whose solution gives the probability mass function (pmf) of the considered process. In the present work, we start with the same three axioms stated in \cite{Feller1} with a small relaxation: we allow the function $\lambda_r$ to depend not only on the number of failures $r$ but also on the elapsed time $t$. This allows to model a wider class of processes with a more general growth pattern. 
	
	By following the same steps, a system of infinite recursive ODEs can be derived from those postulates. For the sake of simplicity, we take this system to be our starting definition. Moreover, we will assume an initial population $N(0)=0$, which is indeed the case in most real scenarios of application.
	
	\begin{definition}[Birth process]\label{def_general_birth_process}
		Let $P_r(t), \; r\in\mathbb{Z}$, denote the probability of the system being in state $r$ at time $t$, and let $0$ be the initial population, i.e:
		\begin{align}
			P_0(0) &= 1,\\
			P_r(0) &= 0 \quad \forall r > 0.
		\end{align}
		Then, for any $t > 0, r \geq 0$ the probability mass function of the process, which we call \textbf{birth process}, is given by:
		\begin{align}
			P'_r(t) &= -\lambda_r(t)\; P_r(t) + \lambda_{r-1}(t)\; P_{r-1}(t), \label{eq_def_gbp}\\
			P'_0(t) &= -\lambda_0(t)\; P_0(t), \label{eq_def_gbp_0}
		\end{align} 
		where $\lambda_r(t)$ is a function, called \textbf{intensity function} or \textbf{failure rate}, that is allowed to depend on both $r$ and $t$.
		For $r < 0$ and any $t$, we have $P_r(t) = 0$.  
	\end{definition}

	Existence and uniqueness of the solution follows directly from the theory of differential equations \cite{coddington1955theory} only requiring the intensity $\lambda_r(t)$ to be a continuous function of $t$. The solution can be expressed explicitly as a system of integral equations (see \cite{sendova}), which may be easier to solve in some cases:
	\begin{align}
		P_r(t) &= \int_0^t \left( e^{-\int_x^t \lambda_r(u)du } \right) \lambda_{r-1}(x)P_{r-1}(x)dx,\\
		P_0(t) &= e^{-\int_0^t \lambda_0(u)du} \label{exponential_wt}. 
	\end{align}

	As it is mentioned in (\cite{grandell1997mixed}, p. 60), in order that $P_r(t)$ are well-defined, it is a sufficient condition that 
	\begin{equation}
		\sum_{r=0}^\infty \max_{0\leq s < t} \lambda_r^{-1}(s) = \infty.
	\end{equation}

	\textit{Remark:} \cref{exponential_wt} is usually known as \textit{exponential waiting time}.

	\subsubsection{Process classification}
	
	Most known point processes can be obtained as special cases of the birth process, by imposing different restrictions to the intensity function. We highlight four cases:
	\begin{itemize}
		\item Non-homogeneous birth processes (NHBP): these are birth processes whose failure rate $\lambda_k(t)$ depends on both $k$ and $t$. The dependence of the intensity on the elapsed time $t$ implies non-homogeneity. The contagion models studied in the literature  \cite{konno2010exact, klugman2013loss} fall into this category.
		\item Homogeneous birth processes: these are  described by an intensity $\lambda_r(t) = \lambda_r$ that does not depend on $t$. They are also called pure birth processes \cite{Feller1}. 
		\item Non-homogeneous Poisson processes (NHPP): a well-known generalization of the regular Poisson process, where the intensity $\lambda_r(t) = \lambda(t)$ does not depend on $r$ but is not constant over $t$.
		\item Ordinary Poisson processes: the intensity $\lambda_r(t) = \lambda$ is a constant, i.e., depends on neither $r$ nor $t$.
	\end{itemize}

	All the processes described here are Markov point  processes that fall under \cref{def_general_birth_process}. Non-homogeneous Poisson processes also present the stronger property of having independent increments \cite{klugman2013loss}. 

\begin{figure}[!h]
		\includegraphics[width=0.8\textwidth]{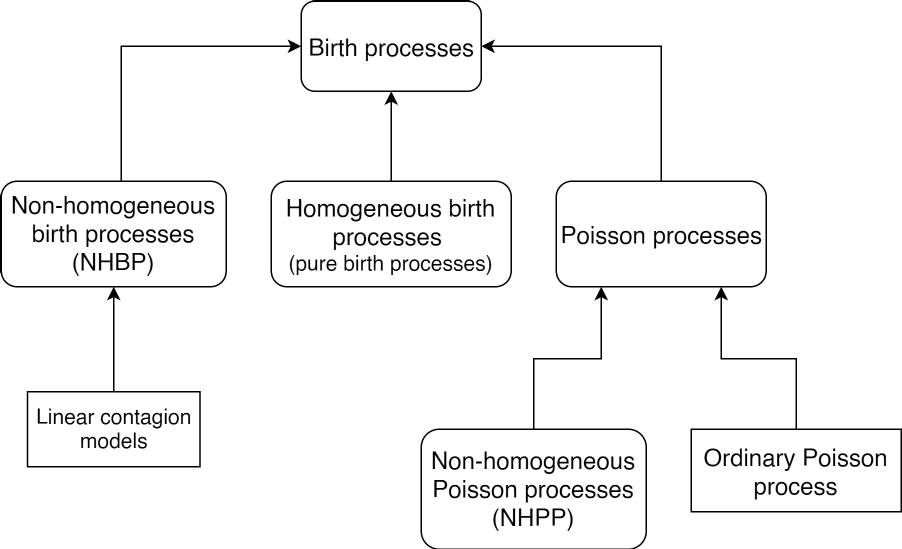}
		\caption{Classification of birth processes}
		\label{fig_clasif_processes}
	\end{figure}

	The classification is depicted in \cref{fig_clasif_processes}.
	
	\subsubsection{Probability mass function}
	
	Although a general expression for the pmf of a  birth process is not available, there exist several particular cases (including the ones we require here) where it is possible to obtain one. 
	
	\begin{enumerate}
		\item Non-homogeneous Poisson processes: if $\lambda_r(t) = \lambda(t)$, then the pmf \cite{klugman2013loss} is given by:
		\begin{equation}\label{eq_p_nhpp}
			P_r(t) = \frac{\mu(t)^r}{r!}e^{-\mu(t)},
		\end{equation} 
		with $\mu(t) = \int_0^t \lambda(x)dx$. \cref{eq_p_nhpp} also gives the correct pmf expression of the ordinary Poisson process, obtained by making $\lambda(t)$ constant over $t$.
		\item Linear contagion processes: these are NHBPs whose intensities can be expressed as
		\begin{equation}\label{linear_contagion}
			\lambda_r(t) = (mr + b)f(t), \quad m\neq 0,
		\end{equation} where $f(t)$ is any continuous function and $m\neq 0$ and $b$ are real constants. In this case, the pmf can be expressed as:
		\begin{equation}\label{eq_p_linear_contagion}
			P_r(t) = \frac{\Gamma\left( r + \frac{b}{m} \right)}{\Gamma \left( \frac{b}{m} \right)} \frac{e^{-\mu_r(t)}}{r!} \left( e^{\Delta(t)} - 1  \right)^r,
		\end{equation}
		where $\mu_r(t) = \int_0^t \lambda_r(x)dx$ and $\Delta(t) = \mu_{r+1}(t) - \mu_r(t)$ independent of $r$. 
		\textit{Remark}: $\Delta(t)$ not depending on $r$ is equivalent to $\lambda_r(t)$ having the linear form described in \cref{linear_contagion}. The proof can be found in \cite{klugman2013loss}; we have also shown in \cite{CSFEpidemiologia} that a particular case of this expression holds for the contagion process described there. The well-known Yule and Polya processes \cite{Feller1,Yule,lundberg1964random} are also special cases of these, and their pmf can be obtained using \cref{eq_p_linear_contagion}.
	\end{enumerate}

	\subsection{The mean value function}

	The mean value function (mvf) of a birth process, let's call it $M(t)$, is extremely useful for the applications and, in most cases, quite easy to find. The mvf is defined as:
	\begin{equation}\label{def_mvf}
		M(t) = \sum_{r=0}^{\infty} r \; P_r(t).
	\end{equation}
	The following theorem gives a useful tool to calculate $M(t)$.
	\begin{theorem}\label{teo_mvf}
		Let $M(t)$ be the mean value function (\cref{def_mvf}). $M(t)$ is the solution of the differential equation
		\begin{equation}
			M'(t) = \sum_{r=0}^\infty \lambda_r(t)\; P_r(t), 
		\end{equation}
		provided that $M(t)$ is finite and $\lim\limits_{r\to\infty} r\; \lambda_r(t)\; P_r(t)=0$.
	\end{theorem}

	\begin{proof}
		Multiplying \cref{eq_def_gbp} by $r$ and summing up to $R$ yields:
		\begin{align}
			\sum_{r=0}^R r\; P'_r(t) &= -\sum_{r=0}^R r\; \lambda_r(t)\; P_r(t) + \sum_{r=0}^R r\; \lambda_{r-1}(t)\; P_{r-1}(t) \\
			&= -\sum_{r=0}^R r\; \lambda_r(t)\; P_r(t) + \sum_{r=0}^{R-1} (r+1)\; \lambda_{r}(t)\; P_{r}(t) \\
			&= -R\; \lambda_R(t)\; P_R(t) + \sum_{r=0}^{R-1} \lambda_{r}(t)\; P_{r}(t).
		\end{align}
	
		Letting $R\to\infty$, the first term of the right member vanishes and we get:
		\begin{equation}\label{eq_proof_mvf_dif_equation}
			 \sum_{r=0}^\infty r\; P'_r(t) = \sum_{r=0}^\infty \lambda_{r}(t)\; P_{r}(t).
		\end{equation}
		The theorem will be proved if the derivative $P'_r(t)$ can be extracted from the sum, i.e, if:
		\begin{equation}\label{eq_2_proof_mvf_dif_equation}
			\sum_{r=0}^\infty r\; \frac{\mathrm{d}P(t)}{\mathrm{d}t} = \frac{\mathrm{d}}{\mathrm{d}t} \sum_{r=0}^\infty r\; P_r(t).
		\end{equation}
	
		Although this is not necessarily true, the condition is clearly met if the series in the right member of \cref{eq_2_proof_mvf_dif_equation} converges, this is, if the mean value $M(t)$ is finite for every finite $t$, as assumed. Hence \cref{teo_mvf} follows.
	\end{proof}

	It is straightforward to check that linear contagion processes and NHPPs satisfy $\lim\limits_{r\to\infty} r\; \lambda_r(t)\; P_r(t)=0$. Finiteness of $M(t)$ is a little trickier, but in both cases it can be shown by using D'Alembert's criterion for convergence of series. Hence the conditions of \cref{teo_mvf} hold in all the cases of our interest.
	
	\subsection{Mean time between failures}
	
	In this section we will analyze birth processes on the time domain. 
	
	\subsubsection{Time domain analysis}
	
	We begin by defining random variables corresponding to failure times.  
	\begin{definition}[Time to the $k$-th failure]\label{def_Tk}
		Let $N(t)$ be a birth process with intensity $\lambda_r(t)$ and probability $P_r(t)$. Let $k$ be any positive integer. Then we define for every $k$ the r.v. $T_k > 0$ to be the exact time in which the $k$-th failure is detected. Moreover, we note as $\Omega_k$ the event $T_k < \infty$, i.e., $P(\Omega_k)$ is the probability of the $k$-th failure to be detected at any finite time.  
	\end{definition}
  
	The distribution of the $T_k$ variables can be derived from the process $N(t)$. We state \cref{teo_f_k}.
	
	\begin{theorem}\label{teo_f_k}
		Let $T_k$ be a r.v. defined as in \cref{def_Tk}. If the $k-$th failure is detected within a finite time, then that time has a (conditional) distribution function given by:
		\begin{equation}\label{eq_F_k}
			F_k(t | \Omega_k) = \frac{1 - \sum_{r=0}^{k-1} P_r(t)}{1 - \sum_{r=0}^{k-1} \lim\limits_{x\to\infty} P_r(x)}.
		\end{equation}
	\end{theorem}

	The reason to compute the conditional cdf instead of the marginal is because those probabilities are not defined if $\Omega_k^c$ occurs, i.e., if the failure is never detected at any finite time. This consideration allows a clean definition of the probabilities without any technical difficulties.
	
	\begin{proof}
		By the definiton of conditional probability:
		\begin{equation}\label{eq_conditional_pr}
			F_k(t|\Omega_k) = \frac{P(T_k < t,\; \Omega_k)}{P(\Omega_k)} = \frac{P(T_k < t,\; T_k < \infty)}{P(T_k < \infty)} = \frac{P(T_k < t)}{P(T_k < \infty)}.
		\end{equation}
		The numerator of \cref{eq_conditional_pr} can be written as $1 - P(T_k > t)$, and this last probability equals that of having $k$ or less failures at time $t$. Hence the numerator is proved. 
		
		The denominator can be calculated as:
		\begin{align}
			P(T_k<\infty)&=\lim\limits_{t\rightarrow \infty} P(T_k <t) = \lim\limits_{t\rightarrow \infty} \left( 1-P(T_k \geq t) \right) \\
			&= 1-\sum_{r=0}^{k-1} \lim\limits_{t\rightarrow \infty} P_r(t)\label{lim_P_omega},
		\end{align}
		
		which completes the proof.
	\end{proof}

	In the particular case of the NHPPs, the denominator of \cref{eq_F_k} has a simple expression. We show this in the following theorem. 
	\begin{theorem}\label{teo_O_k_NHPP}
		Let $\Omega_k$ be defined as in \cref{def_Tk}. For NHPPs:
		\begin{equation}\label{O_k_NHPP}
			P(\Omega_k) = \frac{\gamma(L,k)}{(k-1)!}, \qquad L=\lim\limits_{t\rightarrow \infty} \mu(t),
		\end{equation}
		where $\gamma(L,k)$ is the incomplete Gamma function:
		\begin{equation}
			\gamma(L,k)=\int_0^L z^{k-1}e^{-z}dz.
		\end{equation}
	\end{theorem}
	
	\begin{proof}
		We are going to demonstrate this by induction. Since we already know that in a NHPP the pmf is given by \cref{eq_p_nhpp}, we can combine it with \cref{lim_P_omega} to obtain an explicit $P(\Omega_k)$. Thus we have to prove the following equation:
		\begin{equation}
			1-e^{-L}\sum_{r=0}^{k-1} \frac{L^r}{r!}= \frac{\gamma (L,k)}{(k-1)!}.
		\end{equation}
		It is immediate that for $k=1$, $\gamma(L,1)=\int_0^L e^{-z}dz=1-e^{-L}$. 
		For the inductive step we assume that \cref{O_k_NHPP} is valid for $k$ and check that then it holds for $k+1$:
		\begin{equation}
			\frac{\gamma(L,k+1)}{k!}=\frac{1}{k!}\int_0^L z^k e^{-z}dz.
		\end{equation}
		Integrating by parts yields:
		\begin{align}
			\frac{\gamma(L,k+1)}{k!}&=\frac{1}{k!}\Big(-L^ke^{-L}+k \int_0^L z^{k-1}e^{-z}dz \Big)\\
			&=\frac{-L^ke^{-L}}{k!} + \frac{\int_0^L z^{k-1}e^{-z}dz}{(k-1)!} \\
			&=\frac{-L^ke^{-L}}{k!} + \frac{\gamma(L,k)}{(k-1)!}. 
		\end{align}
		Finally, by applying the inductive hypothesis we get:
		\begin{align}
			\frac{\gamma(L,k+1)}{k!}&=\frac{-L^k e^{-L}}{k!} +1-e^{-L}\sum_{r=0}^{k-1} \frac{L^r}{r!}\\
			&=1-e^{-L}\sum_{r=0}^{k} \frac{L^r}{r!},
		\end{align}	
		as we intended to prove.
	\end{proof}

	Differentiating \cref{eq_F_k} and replacing \cref{eq_def_gbp} yields a more convenient expression for the pdf:
	\begin{equation}
		\label{eq_f_k}
		f_k(t | \Omega_k) = \frac{\mathrm{d}F_k(t | \Omega_k) }{\mathrm{d}t} = \frac{\lambda_{k-1}(t)P_{k-1}(t)}{1 - \sum_{r=0}^{k-1} \lim\limits_{x\to\infty} P_r(x)}.
	\end{equation}
	
	In the special case of NHPPs, by \cref{teo_O_k_NHPP} we can write the pdf of $T_k$ as follows:
	\begin{equation}
		\label{eq_f_k_NHPP}
		f_k(t | \Omega_k) = \frac{ (k-1)!\lambda(t)P_{k-1}(t)}{\gamma(L,k)},
	\end{equation}
	with $\gamma$ and $L$ as in \cref{teo_O_k_NHPP}. By substituting \cref{eq_p_nhpp} we get the known expression \cite{Pham:2010:SSR:1965277}:
	\begin{equation}
			f_k(t | \Omega_k) = \frac{\lambda(t)[\mu(t)]^k}{\gamma(L,k)}e^{-\mu(t)}.
	\end{equation}

	In \cite{CSFEpidemiologia} we derive another distribution of interest, that of $T_k$ conditioned to knowing $s=T_{k-1}$ (\cref{eq_fk_condicional}).
	\begin{equation}\label{eq_fk_condicional}
		f_k(t | T_{k-1}=s, \Omega_k) = \frac{\lambda_{k-1}(t)e^{-\int_s^t \lambda_{k-1}(x)dx}}{1 - e^{-\int_s^\infty \lambda_{k-1}(x)dx}}.
	\end{equation}

	The denominator of \cref{eq_F_k,eq_f_k} takes the value 1 when there is zero probability of the failure $k$ never being detected. Some cases, like that of the ordinary Poisson process and the contagion model described in \cref{sec_models}, present this property. The following theorem provides a way to identify those in all the cases concerning this work.
	\begin{theorem}\label{teo_condicion_mu}
		Consider a linear contagion process with $\lambda_k(t)$ of the form specified by \cref{linear_contagion}. Let $T_k$ be defined as in \cref{def_Tk}. For $P(\Omega_k) = P(T_k < \infty)$ to equal one it is necessary and sufficient that
		\begin{equation}
			\mu_r(\infty) = \lim\limits_{t\to\infty} \int_0^t \lambda_r(x)dx = \infty
		\end{equation}
		for all $r < k$.
	\end{theorem}

	\begin{proof}
		 Recalling the definition of  $\Delta(t)$:
		\begin{align}
			\Delta(t) &= \mu_{r + 1}(t) - \mu_{r}(t) \\
			&= \int_0^t (mr + m + b) f(x)dx - \int_0^t (mr+b)f(x)dx \\
			&= m\int_0^t f(x)dx \\
			&= \frac{m}{mr+b}\mu_r(t),
		\end{align}
		it follows from (\ref{eq_p_linear_contagion}) that:
		\begin{align}
			P_r(t) &= \frac{\Gamma\left( r + \frac{b}{m} \right)}{\Gamma\left( \frac{b}{m} \right)} \frac{e^{-\mu_r(t)}}{r!} \left( e^{\Delta(t)} - 1  \right)^r \\
			&= \frac{\Gamma\left( r + \frac{b}{m} \right)}{\Gamma\left( \frac{b}{m} \right)} \frac{e^{-\mu_r(t)}}{r!} e^{\Delta(t)r} \left( 1-e^{-\Delta(t)} \right)^r  \\
			&= \frac{\Gamma\left( r + \frac{b}{m} \right)}{\Gamma\left( \frac{b}{m} \right)} \frac{e^{-\mu_r(t)\left(1-\frac{mr}{mr+b}\right)}}{r!} \left( 1-e^{-\Delta(t)} \right)^r.
		\end{align}
		If $\mu_r(\infty) = \infty$, then the exponentials $e^{-\mu_r(t)\left(1-\frac{mr}{mr+b}\right)}$ and  $e^{-\Delta(t)}$ tend to zero. Hence $P_r(t)$ has limit zero for every $r < k$, which implies that $\sum_{r=0}^{k-1} P_r(t)$ has limit zero. Thus the condition is sufficient.
		
		To see that it is necessary, we assume $\mu_r(\infty) < \infty$. Then $e^{-\mu_r(t)}$ and $e^{-\Delta(t)}$ can never have limit zero, which implies $\sum_{r=0}^{k-1} P_r(t)$ has a positive limit, a contradiction. This proves \cref{teo_condicion_mu}. 
		
		\textit{Remark:} the necessary condition holds even if not every $\mu_r(\infty) < \infty$; it suffices that at least one of them accomplishes it in order for the sum to have nonzero limit.  
	\end{proof}

	A similar result holds for the non-homogeneous Poisson processes as shown below:
	\begin{theorem}
		\label{teo_condicion_mu_nhpp}
		Consider a NHPP with intensity $\lambda(t)$. Let $T_k$ be defined as in \cref{def_Tk}. For $P(\Omega_k) = P(T_k < \infty)$ to equal one for every $k$ it is necessary and sufficient that
		\begin{equation}
			\mu(\infty) = \lim\limits_{t\to\infty} \int_0^t \lambda(x)dx = \infty.
		\end{equation}
	\end{theorem} 

	The proof follows the same steps as that of \cref{teo_condicion_mu}, taking into account that the process pmf is now given by \cref{eq_p_nhpp} instead of \cref{eq_p_linear_contagion}.
	
	\subsubsection{MTTF and MTBF}
	
	Two mean values of interest can be calculated: the \textit{mean time to failure} (MTTF) and the \textit{mean time between failures} (MTBF). Study of the latter is the main objective of this work.
	
	To begin with, we shall state formal definitions for both.
	
	\begin{definition}[MTTF and MTBF]\label{def_mttf_mtbf}
		Let $T_k$ be defined as in \cref{def_Tk}. The mean time to the $k-$th failure is the mean value:
		\begin{equation}\label{eq_def_mttf}
			MTTF_k = E[T_k].
		\end{equation} 
		Let $X_k = T_k - T_{k-1}$ for any $k > 1$. The mean time between the $k-1$ and $k$-th failures is the mean value:
		\begin{equation}\label{eq_def_mtbf}
			MTBF_{k-1, k} = E[X_k],
		\end{equation}
		provided that $MTTF_k$ and $MTTF_{k-1}$ are both finite.
	\end{definition}

	Calculation of the MTTF is straightforward just by taking the expectation of $T_k$, since in practical cases its pdf is known (\cref{eq_f_k}). A more convenient method can be used to obtain the MTBF of processes with $P(\Omega_k) = 1$ for all $k$. 
	\begin{theorem}\label{teo_mtbf_alt}
			Let $T_k$ be defined as in \cref{def_Tk}. If $\lim\limits_{t\to\infty} t\; P_k(t) = 0$ and $P(\Omega_k)=1$ for all $k > 1$, then we have:
			\begin{equation}\label{formula_mttf_alt}
				MTBF_{k-1,k} = \int_0^\infty P_k(t)dt, \qquad k>1.
			\end{equation}
	\end{theorem}

	\begin{proof}
		We begin by writing the definition of $MTTF_{k-1}$ and replacing \cref{eq_def_gbp}:
		\begin{align}
			MTTF_{k-1} &= \frac{1}{P(\Omega_{k-1})} \int_0^\infty t\;\lambda_{k-2}(t)\;P_{k-2}(t)dt \\
			P(\Omega_{k-1})MTTF_{k-1} &= \int_0^\infty t\left[ P'_{k-1}(t) + \lambda_{k-1}(t)\; P_{k-1}(t) \right]dt \\
			P(\Omega_{k-1})MTTF_{k-1} &= \int_0^\infty t \; P'_k(t)dt + P(\Omega_k)MTTF_k
		\end{align}
		Since by hypothesis $P(\Omega_k)$ and $P(\Omega_{k-1})$ are 1:
		\begin{equation}
			MTBF_{k-1,k} = MTTF_k - MTTF_{k-1} = \int_0^\infty t\; P'_k(t)dt.
		\end{equation}
		
		Integrating by parts and using $\lim\limits_{t\to\infty} t\; P_k(t) = 0$ yields \cref{formula_mttf_alt}.
	\end{proof}

	To finalize this section, we will study an alternative definition for the MTBF that is, in practice, much easier to obtain. Since most software failure reports records the time every failure is detected, this information can be used to enhance the calculation by conditioning. Suppose we know the $k-1$-th failure was detected at time $s$; we wish to calculate the model's prediction for the time up to the next failure, which is $MTBF_{k-1, k}$, using our knowledge of $T_{k-1}=s$. This motivates the following definition.
	
	\begin{definition}[Conditional MTBF]\label{def_conditional_mtbf}
		Let $T_k$ be defined as in \cref{def_Tk}. Assume that $s = T_{k-1}$ is known. Then, we define the conditional mean time between the $k-1$ and $k$-th failures as:
		\begin{equation}
			MTBF^{\ast}_{k-1,k}(s,k) = E[T_k|T_{k-1}=s] - s, 
		\end{equation}
		where the expectation shall be computed by \cref{eq_fk_condicional}.
	\end{definition}
	
	This definition has three main advantages. First, since we are using additional information, the resulting $MTBF^\ast$ is a better estimate of the real data. Second, expectation of the conditional density given by \cref{eq_fk_condicional} is much easier to calculate than that of the marginal density (\cref{eq_f_k}). And last, \cref{def_conditional_mtbf} can be used in cases where the MTTF is infinite, which implies that the regular MTBF cannot be calculated by \cref{def_mttf_mtbf}. Moreover, the denominator of \cref{eq_fk_condicional}, which is the probability of $\Omega_k$ conditioned to the knowledge of $T_{k-1}$, is also easy to calculate.

\section{Application}

	In this section, we will use different models based on birth processes to study the failure detection process on two software projects developed in recent years under agile methodologies. The datasets were gently provided by the agile software development company Grupo Esfera S.A. \cite{esfera}. We begin by presenting the models.

\subsection{Models considered}\label{sec_models}

	\cref{tab_models} contains a non exhaustive list of software reliability models based on birth processes studied in the literature. Except for the linear contagion model, the rest are traditional models based on NHPPs. The Musa-Okumoto and Goel-Okumoto models are most suitable to model a reliability growth behaviour, but evidence presented in \cite{penabarrazaanand} suggest that they cannot model an increasing failure rate. In that work, we have shown that due to this, those models are not good for describing agile projects like the ones we analyze here. The Delayed S-shaped and Logistic are NHPPs with the capacity of modeling datasets with two different stages of growth (as it can be seen in \cite{penabarrazaanand}, their mean value curves have an inflection point). Finally, the linear contagion model we consider is obtained as a special case of \cref{linear_contagion}. An interesting feature of this model is that it is able to model both reliability growth and increasing failure rate behaviours.
	
\begin{table}[!h]
    \centering
    \large
    \caption{Software reliability models based on birth processes}
    \label{tab_models}
    \vspace{0.5em} 
    \begin{tabularx}{\textwidth}{>{\raggedleft\arraybackslash}X | >{\centering\arraybackslash}X}
        \hline 
        Model & $\lambda_r(t)$ \\
        \hline\hline
        Musa-Okumoto & $\frac{ab}{1 + bt}, \quad a, b > 0$ \\
        \hline
        Goel-Okumoto & $abe^{-bt}, \quad a, b > 0$ \\
        \hline
        Delayed S-shaped & $ab^2 te^{-bt}, \quad a, b > 0$ \\
        \hline
        Logistic & $\frac{abe^{-b(t - c)}}{[1 + e^{-b(t-c)}]^2}, \quad a, b, c > 0$ \\
        \hline
        Linear contagion & $\rho\frac{1 + \frac{\gamma}{\rho}r}{1 + \rho t}, \quad \gamma, \rho > 0$ \\
        \hline
    \end{tabularx}
\end{table}

	More information about contexts in which some models may perform better the rest can be obtained by applying \cref{teo_condicion_mu,teo_condicion_mu_nhpp}. The Delayed S-shaped and Logistic models have nonzero probability of the population arriving at a finite stationary state, while the Musa-Okumoto and linear contagion models do not: every finite $k$-th failure has probability of being detected, which suggests that these processes are better suited to model populations that may grow with no limits. This is indeed the case of failures detected in an Agile project: since new code is constantly added, the failure population continues to grow non-stop. Processes like the Delayed S-shaped and the Logistic are more suitable to model populations with an upper limit; however, contagion model has proved to perform quite well in the early stages of growth within a finite population, i.e., in the segment of the time axis where the upper limit is much larger than the process state (see \cite{CSFEpidemiologia}).

\subsection{Experiments}

	We consider two datasets corresponding to software projects developed under Agile. The application of a model consists mainly on the calibration of the parameters, which can be done by different methods. In this work, the calibration is performed by a nonlinear least-squares fitting, using the Levenberg-Marquardt algorithm. A comparison of calibration methods is presented in \cite{penabarrazaanand}, where we found least-squares to perform better than maximum likelihood. 
	Using this estimation procedure, the model is, of course, sensitive to the quality of the data. The presence of outliers for example results on a jump in the cumulative values, altering this way the outcomes. The same situation arises when the measuring methodology changes "on the run"; in this cases, a multistage analysis might be a better approach, see for example \cite{8701285}. This kind of "sudden changes" in the behavior of the data is detected by the model, typically producing less accurate fits (which can be quantified, for example, by a lower R$^2$ coefficient), and should be carefully treated in order to prevent wrong results.
	
	The MTBF calculation is done by \cref{def_mttf_mtbf} for the NHPP-based models. The linear contagion model requires a more careful analysis. By replacing \cref{eq_p_linear_contagion} in \cref{eq_def_mttf} it can be seen that $E[T_k]$ diverges, which implies the MTTF cannot be defined. This does not imply that the MTBFs are also infinite, but it makes impossible to calculate them via \cref{def_mttf_mtbf}. Hence we turn to \cref{def_conditional_mtbf} and calculate the conditional MTBF, which yields: 
	\begin{equation}
		MTBF^\ast_{k-1,k}(s,k) = \frac{1 + \rho s}{\gamma(k-1)}. 
	\end{equation} 
	A complete derivation of this result is done in \cite{CSFEpidemiologia}.

\subsubsection{Project \#1}

\begin{figure}[!h]
		\includegraphics[width=0.8\textwidth]{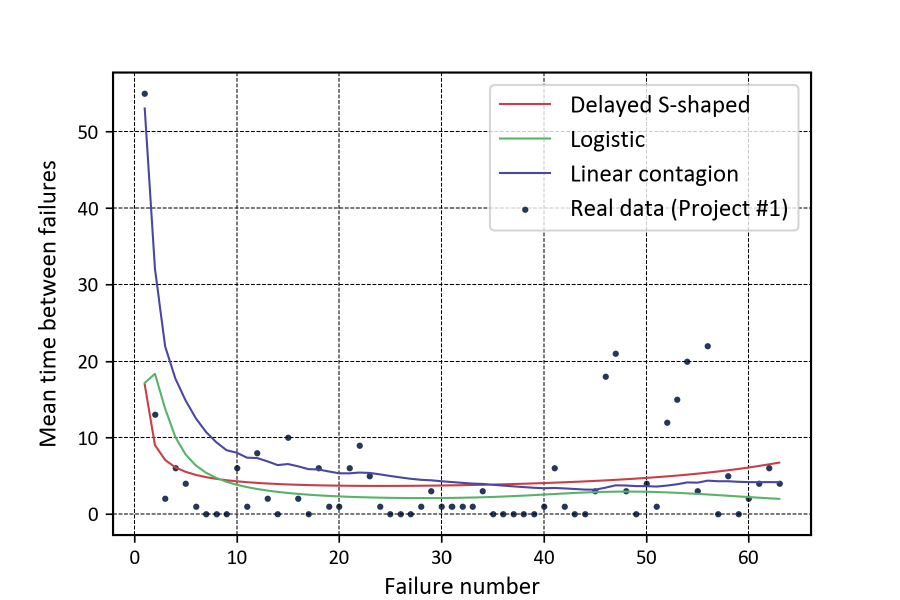}
		\caption{Project \#1 MTBF curves}
		\label{fig_project_1}
	\end{figure}

	\cref{fig_project_1} depicts the resulting interpolated MTBF curves of the three models and the real data from the failure report. The $x$ axis shows the failure number $k$, and $y$ axis shows $MTBF_{k,k+1}$. MTBF is measured in days. The data show that most of the failures were detected very close in time (note that there is a region with a high density of points in the lower part of the plot). This makes the comparison more difficult, since the three models follow the data very similarly. There are some outlier values between the 40 and 60th failures, which none of the models detect. Slight differences can be appreciated in the shape of the curves. The Delayed S-shaped curve is quite smooth, with a perfectly defined bathtub shape. The Logistic curve exhibits a sharp peak at the early part, after which it has several increasing and decreasing stages. The conditional MTBF curve of the contagion model is not as smooth as the others: a clear bathtub envelope can be appreciated as well as some high frequency noise. It is interesting to observe that this curve is the only one to detect the first datum, which is a rather heavy outlier: this is due to the fact that its information is included in the conditioning, one of the main advantages of this method.  

\subsubsection{Project \#2}

\begin{figure}[!h]
		\includegraphics[width=0.8\textwidth]{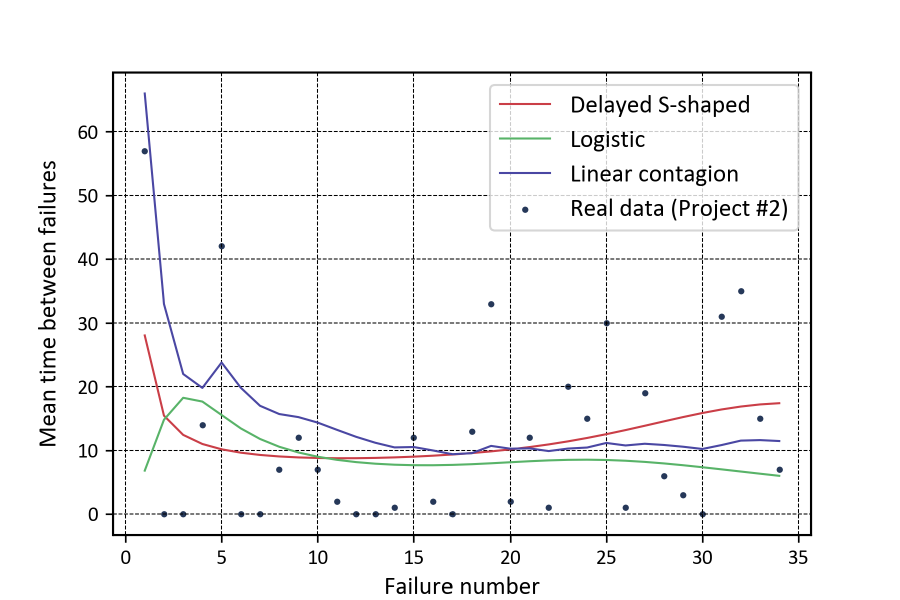}
		\caption{Project \#2 MTBF curves}
		\label{fig_project_2}
	\end{figure}

	MTBF curves for this project are shown in \cref{fig_project_2}. MTBF is measured in days. The Delayed S-shaped and Logistic curves show the same behaviour as in Project \#1, the first with a smooth bathtub shape and the second with successive increasing and decreasing stages and a peak at the beginning. As can be deduced by the low density of points between the 1 and 10th days, failures are quite scattered in time, which may suggest that this project is reaching a reliability growth stage. This behaviour is only captured by the Delayed S-shaped model. As in Project \#1, the linear contagion model (conditional MTBF) is the only one to detect the early outliers.

\subsection{Discussion}

	Results suggest that the behaviour of the MTBF in agile software projects does not differ much from that of hardware reliability studies. An intensity failure rate is observed at the beginning, represented by a decreasing stage in the MTBF; this is typical of agile projects, as pointed by previous works on the topic (see \cite{penabarrazaanand}). The models differ, since both Delayed S-shaped and Logistic curves present an increasing stage in the late part of the dataset, which corresponds to a reliability growth stage in the failure detection process, while the contagion model places the whole dataset in the increasing failure rate phase. This was already observed in \cite{penabarrazaanand}, and it is consequence of the absence of an inflection point in the model's mean value curve. Logistic model also presents some minor differences such as the initial peak or the presence of more than two increase-decrease phases. However, the differences are appreciated only in the shape, since numerical values do not differ much at all, remaining most of the time within a short range not far from the real data. 

\section{Conclusions}

A summary on the MTTF and MTBF metrics as obtained from several Software Reliability models described under the framework of non-homogeneous birth processes was presented. A complete self-contained mathematical background in order to show how those metrics are obtained from the non-homogeneous birth processes theory was developed. The well known Software Reliability models based on non-homogeneous Poisson processes and the recently proposed contagion models were analyzed. A comparison of the performance of applying those models to real datasets was also shown and discussed.  

\section*{Acknowledgment}

The authors thank Universidad Nacional de Tres de Febrero for support under grant no. 32/19 80120190100010TF.

\bibliography{biblio}

@book{Feller1,
    author = {Feller, W.},
    day = {01},
    edition = {3rd},
    howpublished = {Hardcover},
    isbn = {0471257087},
    month = jan,
    publisher = {Wiley},
    title = {An Introduction to Probability Theory and Its Applications, Vol. 1},
    year = {1968}
}

@book{barlow,
	title={Mathematical Theory of Reliability},
	author={Barlow, R. and Proschan, F.},
	publisher = {Society for Industrial and Applied Mathematics},
	address = {Philadelphia, PA 19104-2688},
	URL = {http://epubs.siam.org/doi/abs/10.1137/1.9781611971194},
	isbn={9781611971194},
	year={1996},
}

@book{Pham:2010:SSR:1965277,
	author = {Pham, H.},
	title = {System Software Reliability},
	year = {2010},
	isbn = {184996971X, 9781849969710},
	edition = {1st},
	publisher = {Springer Publishing Company, Incorporated},
}

@book{lundberg1964random,
	title={On Random Processes and Their Application to Sickness and Accident Statistics},
	author={Lundberg, O.},
	lccn={66091752},
	url={https://books.google.com.ar/books?id=mr4rAAAAYAAJ},
	year={1964},
	publisher={Almqvist \& Wiksells boktryckeri-a.-b.}
}

@Article{CSFEpidemiologia,
  author   = {N.R. Barraza and G. Pena and V. Moreno},
  title    = {A non-homogeneous {Markov} early epidemic growth dynamics model. {Application} to the {SARS-CoV-2} pandemic},
  journal  = {Chaos, Solitons \& Fractals},
  year     = {2020},
  volume   = {139},
  pages    = {110297},
  issn     = {0960-0779},
  doi      = {https://doi.org/10.1016/j.chaos.2020.110297},
  url      = {http://www.sciencedirect.com/science/article/pii/S0960077920306937},
}

@INPROCEEDINGS{7789377,
author={N. R. Barraza},
booktitle={2016 IEEE International Symposium on Software Reliability Engineering Workshops (ISSREW)},
title={Software Reliability Modeled on Contagion},
year={2016},
pages={49-50},
doi={10.1109/ISSREW.2016.22},
month={Oct},}

@article{sendova,
	author = {Sendova, K. and Minkova, L.},
	year = {2019},
	month = {09},
	pages = {1-19},
	title = {Introducing the non-homogeneous compound-birth process},
	journal = {Stochastics},
	doi = {10.1080/17442508.2019.1666132}
}

@incollection{barrazaanand,
  author      = "N. R. Barraza",
  title       = "Software Reliability Modeling for Dynamic Development Environments",
  editor      = "Anand, A. and Ram, M.",
  booktitle   = "Recent Advancements in Software Reliability Assurance",
  publisher   = "CRC Press",
  year        = 2019,
  pages       = "29-37",
  chapter     = 3,
  isbn={9780429776564},
  series={Advances in Mathematics and Engineering},
  url={https://books.google.com.ar/books?id=5p6RDwAAQBAJ},
}

@Book{coddington1955theory,
  title     = {Theory of ordinary differential equations},
  publisher = {McGraw-Hill Education},
  year      = {1955},
  author    = {Coddington, E. and Levinson, N.},
}

@Article{konno2010exact,
  author    = {Konno, Hidetoshi},
  title     = "On the exact solution of a generalized {P{\'o}}lya process",
  journal   = {Advances in mathematical physics},
  year      = {2010},
  volume    = {2010},
  publisher = {Hindawi},
}

@article{Yule,
	ISSN = {02643960},
	URL = {http://www.jstor.org/stable/92117},
	author = {G. Udny Yule},
	journal = {Philosophical Transactions of the Royal Society of London. Series B, Containing Papers of a Biological Character},
	pages = {21--87},
	publisher = {The Royal Society},
	title = {A Mathematical Theory of Evolution, Based on the Conclusions of Dr. J. C. Willis, F.R.S.},
	volume = {213},
	year = {1925}
}

@misc{esfera,
	title = "{Grupo} {Esfera} {S.A.}",
	howpublished = {\url{https://www.grupoesfera.com.ar/}},
	note = {Accessed: 2021-02-22}
}

@Book{grandell1997mixed,
  title     = {Mixed poisson processes},
  publisher = {CRC Press},
  year      = {1997},
  author    = {J. Grandell},
  volume    = {77},
}

@Book{klugman2013loss,
  title     = {Loss models: Further topics},
  publisher = {John Wiley \& Sons},
  year      = {2013},
  author    = {S. A. Klugman and H. H. Panjer and G. E. Willmot},
}

@book{Lyu:1996:HSR:239425,
 editor = {Lyu, M. R.},
 title = {Handbook of Software Reliability Engineering},
 year = {1996},
 isbn = {0-07-039400-8},
 publisher = {McGraw-Hill, Inc.},
 address = {Hightstown, NJ, USA},
}

@incollection{penabarrazaanand,
  author      = "G. Pena and N. R. Barraza",
  title       = "Increasing Failure Rate Software Reliability Models for Agile Projects: 
A Comparative Study",
  editor      = "Adarsh Anand and Mangey Ram",
  booktitle   = "Systems Performance Modeling",
  publisher   = "De Gruyter",
  year        = 2020,
  chapter     = 8,
  isbn={9783110607635},
  series={Applications of Mathematics in Engineering and Information},
  url={https://www.degruyter.com/document/isbn/9783110607635/html},
}

@Article{Deepika,
  author={Deepika and Adarsh Anand and Ompal Singh and P. K. Kapur},
  title={{Three-dimensional wiener process based entropy prediction modelling for OSS}},
  journal={International Journal of System Assurance Engineering and Management},
  year=2021,
  volume={12},
  number={1},
  pages={188-198},
  month={February},
  doi={10.1007/s13198-020-01040-},
  url={https://ideas.repec.org/a/spr/ijsaem/v12y2021i1d10.1007\_s13198-020-01040-4.html}
}

@Article{Shiva,
  author={Tyagi Shiva and Devendra Kumar and Sachin Kumar},
  title={{Open source software: analysis of available reliability models keeping security in the forefront}},
  journal={International Journal of Information Technology},
  year=2019,
  month={February},
  doi={10.1007/s41870-019-00293-y},
  url={https://doi.org/10.1007/s41870-019-00293-y}
}

@INPROCEEDINGS{8701285,
author={N. R. {Barraza}},
booktitle={2019 Amity International Conference on Artificial Intelligence (AICAI)},
title={Software Reliability Analysis of Multistage Projects},
year={2019},
volume={},
number={},
pages={67-73},
doi={10.1109/AICAI.2019.8701285},
ISSN={},
month={Feb},}

\end{document}